\documentclass{nature}
 
\usepackage{graphicx,graphics}
\usepackage{lineno}

\def \CHthp{\ifmmode{\rm CH_3^+}\else{CH$_3^+$}\fi}
\def \CHp{\ifmmode{\rm CH^+}\else{CH$^+$}\fi}
\def \SHp{\ifmmode{\rm SH^+}\else{SH$^+$}\fi}
\def \Cp{\ifmmode{\rm C^+}\else{C$^+$}\fi}
\def \twC{\ifmmode{\rm ^{12}C}\else{$^{12}$C}\fi}
\def \CO{\ifmmode{\rm CO}\else{CO}\fi}
\def \thC{\ifmmode{\rm ^{13}C}\else{$^{13}$C}\fi} 
\def \thCHp{\ifmmode{\rm ^{13}CH^+}\else{$^{13}$CH$^+$}\fi}
\def \twCHp{\ifmmode{\rm ^{12}CH^+}\else{$^{12}$CH$^+$}\fi}
\def \HH{\ifmmode{\rm H_2}\else{H$_2$}\fi}
\def \twCO{\ifmmode{\rm ^{12}CO}\else{$^{12}$CO}\fi}
\def \HCOp{\ifmmode{\rm HCO^+}\else{HCO$^+$}\fi}
\def \HtOp{\ifmmode{\rm H_2O^+}\else{H$_2$O$^+$}\fi}
\def \amm{\ifmmode{\rm NH_3}\else{NH$_3$}\fi}
\def \wat{\ifmmode{\rm H_2O}\else{H$_2$O}\fi}
\newcommand{\msol}{\hbox{\,${\rm M}_\odot$}}

\newcommand{\lsol}{\hbox{\,${\rm L}_\odot$}}
\def\kms{\hbox{\,km\,s$^{-1}$}}
\def\cc{\ifmmode{\rm \,cm^{-3}}\else{\,cm$^{-3}$}\fi}
\def\eccs{\ifmmode{\rm \,erg\,cm^{-3}\,s^{-1}}\else{\,erg\,cm$^{-3}$\,s$^{-1}$}\fi}
\def\es{\ifmmode{\rm \,erg\,s^{-1}}\else{\,erg\,s$^{-1}$}\fi}
\def\gs{\mathrel{\raise0.35ex\hbox{$\scriptstyle >$}\kern-0.6em
\lower0.40ex\hbox{{$\scriptstyle \sim$}}}}
\def\ls{\mathrel{\raise0.35ex\hbox{$\scriptstyle <$}\kern-0.6em
\lower0.40ex\hbox{{$\scriptstyle \sim$}}}}

\def\cq{\ifmmode{\,{\rm cm}^{-2}}\else{\,cm$^{-2}$}\fi}
\title{Large turbulent reservoirs of cold molecular gas 
around high-redshift starburst galaxies }

\author{E.~Falgarone$^1$, M.\,A.~Zwaan$^2$, B.~Godard$^1$,  E.~Bergin$^3$, R.\,J.~Ivison$^{2,4}$,
   P.\,M.~Andreani$^2$,  F.~Bournaud$^5$, R.\,S.~Bussmann$^6$, D.~Elbaz$^5$, A.~Omont$^7$, 
  I.~Oteo$^{4,2}$ \& F.~Walter$^8$}

\begin{document}

\maketitle

\begin{affiliations}
 \item LERMA/LRA, Observatoire de Paris, PSL Research University, CNRS, Sorbonne
Universit\'es, UPMC Universit\'e Paris 06, Ecole normale sup\'erieure, 75005 Paris, France
 \item European Southern Observatory, Karl-Schwarzschild-Strasse 2, 85748 Garching, Germany
 \item  University of Michigan, Ann Arbor, MI, USA
 \item  Institute for Astronomy, University of Edinburgh, Blackford Hill, Edinburgh EH9 3HJ, UK
 \item  CEA/AIM, Saclay, France
 \item Cornell University, Cornell, NY, USA
\item  IAP, CNRS, Sorbonne Universit\'es, UPMC Univiversit\'e Paris 06, 75014 Paris, France
 \item  Max Planck Institute f\"ur Astronomie, Heidelberg, Germany

\end{affiliations}

\begin{abstract}
Starburst galaxies at the peak of cosmic star formation$^1$ 
are among the most extreme star-forming engines in the universe, producing stars over $\sim$ 100~Myr$^2$. 
The star formation rates of these galaxies, which exceed 100 \msol\ per year,
 require large reservoirs of cold molecular gas$^3$ to be delivered to their cores, 
 despite strong feedback from stars or active galactic nuclei$^{4,5}$. Starburst galaxies are therefore
 ideal targets to unravel the critical interplay between this feedback and the growth of a galaxy$^6$.
The methylidyne cation, \CHp, is a most useful molecule for such studies because it
 cannot form in cold gas without supra-thermal energy input, so its presence 
 highlights dissipation of mechanical energy$^{7-9}$ or strong UV irradiation$^{10,11}$.
Here, we report the detection of \CHp($J=$1-0) emission and 
absorption lines in the spectra of six lensed starburst  galaxies$^{12-15}$ at redshifts $z \sim$ 2.5. 
This line has such a high critical density for excitation that it is emitted only 
in very dense ($>10^5$\,\cc)  gas, and is absorbed in low-density gas.
We find that the \CHp\ emission lines, which are broader than $1000$~\kms, originate in 
dense shock waves powered by hot galactic winds. The \CHp\ absorption lines reveal  
highly turbulent reservoirs of cool ($T \sim 100$ K), low-density gas, 
extending far outside  ($> 10$~kpc)  the starburst cores (radii $<1$\,kpc).
 We show that the galactic winds sustain turbulence 
 in the 10~kpc-scale environments of the starburst cores, processing these environments 
 into multi-phase, gravitationally bound reservoirs. However, the mass outflow rates
are found to be insufficient to balance the star formation rates. Another mass input is therefore required for these reservoirs,  
which could be provided by on-going mergers$^{16}$ or cold stream accretion$^{17,18}$.
Our results suggest that galactic feedback, coupled jointly to turbulence and gravity, 
extends the starburst phase instead of quenching it.
  \end{abstract}
   
Using the Atacama Large Millimeter Array (ALMA), we have detected 
\CHp($J$=1--0) lines in high-redshift galaxies:
our sample (Table~1) consists of six particularly bright -- by virtue
of gravitational lensing -- submillimeter-selected galaxies (SMGs), with
specific star-formation rates $\gs 3\times$ those of galaxies
on the main sequence$^{19}$ at $z\sim2$. The SMGs are extremely compact, with half-light radii ($r_{\rm SMG}$)
in the range 0.3--1.2 kpc inferred from lens modeling (Table~1).  
 Their rest-frame 360-$\mu$m dust continuum images and \CHp($J$=1--0) spectra
are displayed in Figs.~1 and 2. 
Only the faintest target, SDP130, is not detected in absorption. 
The peak optical depths at the velocity resolution of 50 \kms\ are large, in the range 0.25--1.2 (see Methods, Table~2).
The absorption lines are broad  (mean linewidth $\Delta  v_{\rm abs} \sim$ 440\kms) and three from five are blue-shifted relative to the \CHp\ emission-line centroid ($v_{\rm em}$, Table~2).  
In the Eyelash, the \CHp\ absorption covers the same velocity range as the OH 119$\mu$m absorption$^{20}$.
Three of the \CHp\ emission lines are broader than 1,000\kms\ (full width at zero intensity, FWZI
$>$ 2,500\kms), far broader than the known CO and \wat\
lines of these galaxies$^{21-23}$ (Extended Data Table 1). 
Unlike the absorption lines, these broad emission lines are all centred 
within $\sim$ 30 \kms\ of the known redshift of the SMGs  and their width is independent of the star formation rate (Tables~1 and 2).   \\
We detect absorption in five targets, out of the six observed.
This high detection rate suggests that the absorbing gas has a quasi-isotropic distribution, a large covering fraction 
on the SMG dust continuum emission and is a feature common to the starburst phase.
The lifetime of the absorbing medium is therefore comparable to 
that of the starburst phase, $t_{\rm SMG} \approx$ 100\,Myr from SMG samples$^{2}$.
The low density gas that is traced by \CHp\ absorption cannot be confined within the starburst cores, because its thermal pressure 
is orders of magnitude below the high pressure in these $z\sim2$ starburst galaxies$^{24}$.  
We ascribe the absorption linewidths to  turbulence (see Methods) 
with mean turbulent velocity $\overline v_{\rm abs}= 0.7 \Delta  v_{\rm abs}$. \\
 \CHp\ column densities are derived from the absorption line profiles (see Methods, Table~2).  
 The \CHp\  molecules must form in the regions of  dissipation of turbulence because, once formed, their lifetime is so short  that they cannot be transported. 
 Their abundance is inferred from the turbulent energy flux that sustains the {\it number} of observed \CHp\ molecules (see Methods). This energy flux 
 depends on the unknown radius $r_{\rm TR}$ of the turbulent reservoirs, being proportional to $ \overline v_{\rm abs}^3/r_{\rm TR}$.
 Importantly,
in the two targets with enough ancillary data on stellar and gas masses (the Cosmic Eyelash$^{24}$ and SDP17b$^{25}$), the  
mean turbulent velocities are equal to the escape velocities $v_{\rm esc}(r) = (2GM_{\rm tot}/r)^{1/2}$ at radii $r=$15.6~kpc and 22.7~kpc respectively. 
Here, $G$ is the gravitational constant and $M_{\rm tot}$ is the sum of the stellar ($M_{*}$) and gas ($M_{\rm gas}$) masses.
The key provision here is that the associated dynamical times, $t_{\rm dyn} =r/ \overline v_{\rm abs}=67$ and 46 Myr respectively, are commensurate 
whereas the stellar masses of these galaxies differ by a factor of 10. 
We thus adopt $t_{\rm TR} \sim 50$\,Myr as an approximate age of the observed turbulent reservoirs, and $r_{\rm TR}=\overline v_{\rm abs}\,t_{\rm TR}$
 for their radius. This radius is a lower limit because no dark matter contribution to the total mass $M_{\rm tot}$ was assumed.
 The estimated radii, in the range $\sim$10--20\,kpc (Table~2), lead to \CHp\ abundances close to those in the Milky Way.
The derived masses of the turbulent reservoirs, $M_{\rm TR}$, assuming a radial mean density distribution (see Methods), are 
$\approx 0.8$--1.4 $\times10^{10}$\msol\ (Table~2). These masses increase the gas mass fraction $M_{\rm gas}/M_{\rm tot}$  by  only 10\% in the Cosmic Eyelash and SDP17b (see Methods). 
The kinetic luminosities of the turbulent reservoirs, 
 $L_{\rm TR} =\frac{1}{2} \,M_{\rm TR} \overline{v}_{\rm abs}^2/t_{\rm TR}$ are in the range 1.2--4.5$ \times10^{9}$\lsol\ (Table 2).\\
The broad \CHp\ emission lines trace gas denser than $\sim 10^5$\cc\ with prodigious velocity dispersions. 
However, the \CHp\ molecules cannot form in shocks faster than $\sim$90\kms, otherwise \HH\ (required to form \CHp)
would be dissociated$^{26}$. 
Instead, they must originate in myriad lower velocity shocks that are sufficiently numerous not to be diluted. 
Irradiated magnetized shocks, propagating at $v_{\rm sh}\sim40\,$\kms\  in dense pre-shock gas, are those most efficient at producing \CHp\ (see Methods). 
The \CHp\ column densities of the shocked gas are high enough to emit lines that are much brighter 
than those detected, allowing for substantial beam dilution.  \\
The \CHp\ spectra provide considerable insight into the physics of these extreme star-forming systems.
Although the sample size is still small, the concomitance of the broad emission and deep absorption lines of \CHp\ suggests that
 they trace coeval processes, i.e. the dense shocks with high velocity dispersion and the dilute turbulent reservoirs are tied together. 
 Yet, the disparity between the emission and absorption linewidths is striking.
The emission linewidths (1100--1400~\kms) in the three sources, the Cosmic Eyelash, G09v1.40 and SDP17b, where they are well determined, 
are very similar whereas their star formation rates span a factor of $\sim$5.  
This finding suggests that these lines trace the fast thermal expansion of hot winds at similar temperatures $T$ in the three sources. 
For outflow velocities $v_{\rm out} \sim \sqrt{3} c_{\rm s}$, where $c_{\rm s} \sim 500\,\kms (\frac{T}{4\times10^7 {\rm K}})^{1/2}$ is the sound velocity of the hot wind, the range 
of observed values, $v_{\rm out}=0.7\,\Delta v_{\rm em}\sim$\,700--10$^3$\,\kms, provides the range $T\sim$ 3--6$\times 10^7$\,K for the hot wind temperature.
This high-velocity outflowing hot wind drives large scale shocks in the dense outskirts of the starburst galaxies. Kinetic energy cascades within these shocks down to the velocity of the shocks that are able to form \CHp, in a process reminiscent to what is observed in the  35\,kpc-long intergalactic shock in the group of galaxies known as Stephan's Quintet$^{27}$. \\
The outflowing wind eventually escapes from each galaxy, generating turbulence in its environment.
It is this transition to turbulence that causes the outflow momentum to change from mostly outward to random and 
allows the gas to be re-captured gravitationally. 
This is why the absorption linewidth in the Eyelash (which is 10 $\times$ less massive than SDP17b) is smaller than in SDP17b.
By preventing a fraction of the outflow from escaping the large-scale galaxy potential well, turbulence makes this fraction available for further star formation$^{28}$. 
Turbulence therefore mitigates the negative feedback of galactic winds on star formation, drawing out the starburst phase.\\
The \CHp\ molecules that are seen in absorption are observed where they form, 
i.e. in the turbulent dissipation sites of a $\sim 100$K dilute molecular phase; this suggests
that the hot outflows also induce phase transitions in the galaxy environments that could be the giant haloes of neutral hydrogen at $10^4$K detected in Ly$\alpha$$^{29}$ at high redshift. 
Importantly, the mass outflow rates that are required to sustain the kinetic luminosity $L_{\rm TR}$  of the turbulent reservoirs closely follow the star formation rates, ${\rm SFR}$ (see Methods): 
$$
 \dot M_{\rm out} = \frac{0.05-0.1}{\eta} \,{\rm SFR} \, (\frac{v_{\rm out}}{800{\rm \,km\,s}^{-1}})^{-2}, 
 $$
 $\eta<1$ being the unknown efficiency of the energy transfer. 
This is so because the ratios $L_{\rm TR}/{\rm SFR}$ 
are observed to be the same within a factor of $\sim$2 for all the sources (Tables~1 and 2).
If mass loss occurs at 
$v_{\rm out} \sim 700$--10$^3$\kms, as suggested by the broad emission lines, then the mass outflow rates are  probably lower than the star formation rates. Therefore, given their masses, the turbulent 
reservoirs that are drained by star formation cannot be fed by outflows only. They must be replenished over the lifetime of the starburst  by other inflows, from tidal streams of on-going major mergers or from cold stream accretion. The diffuse gas reservoirs result
from the mixing of the gas ejected by the starburst-driven outflows with gas present in the wider galaxy environment.\\
By highlighting turbulent dissipation in cool molecular gas, \CHp\ shows that mechanical energy, fed by gravity and galactic feedback, is largely stored in  
turbulence and eventually dissipated at low temperature rather than radiated away by hot gas.  
This turbulent energy storage is observed at two different stages of the feedback process:
in the high-velocity dispersion shocked dense gas close to the starburst cores, 
and very far from them, in the 10kpc-scale massive reservoirs of low-density gas.  
Turbulence appears to be a key process in the triggering and subsequent regulation of star formation.

\noindent
{\bf Received: 26 January 2017; Accepted: 21 June 2017.\\
Published online 14 August 2017.}

\newpage


\begin{figure}
\begin{center}
\includegraphics[angle=0,width=0.8\textwidth]{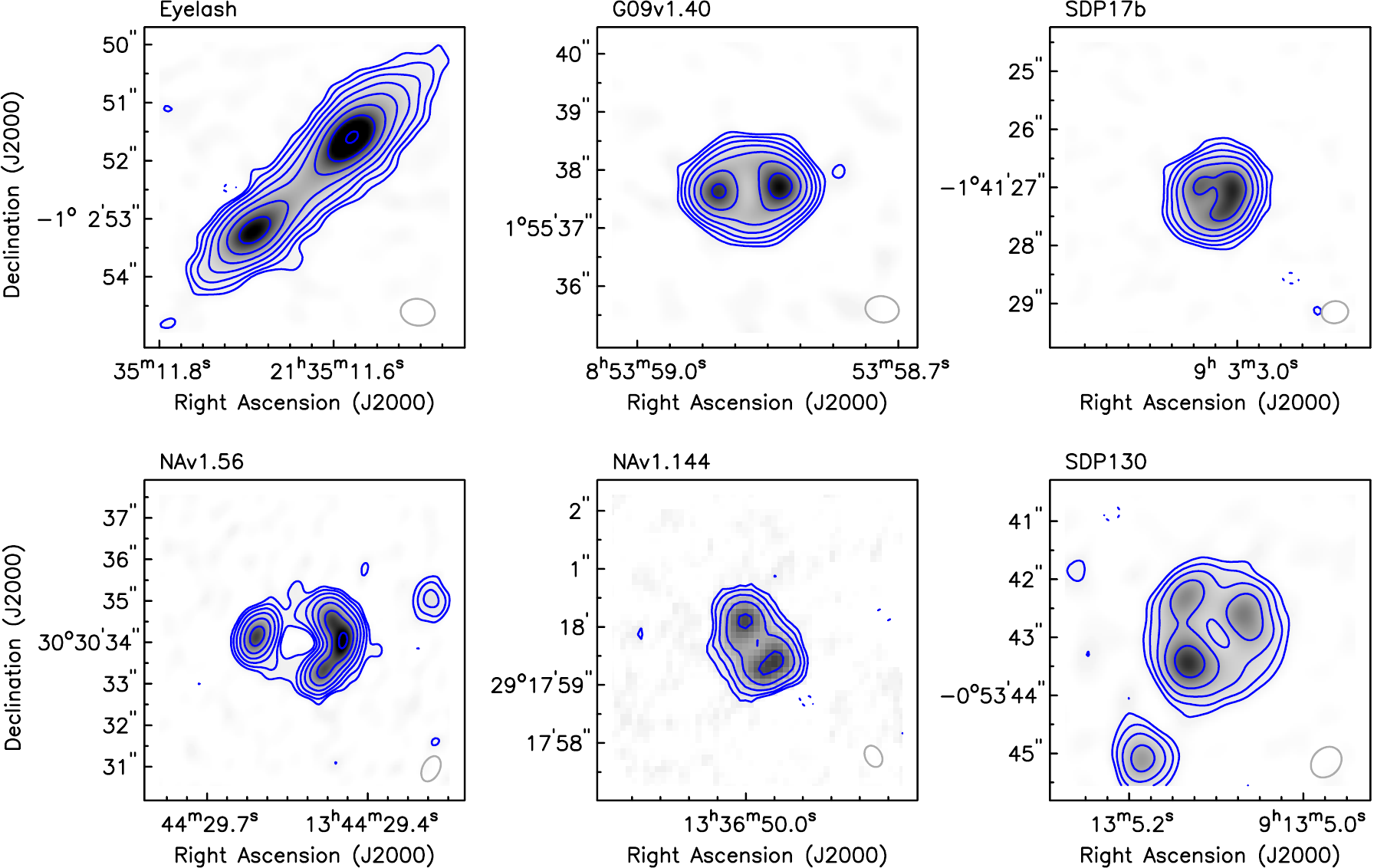}
\end{center}
\caption{{\bf ALMA rest-frame 360-$\mu$m continuum images  of our lensed targets.  }
  Solid contours are in steps of 3$\sigma \times 2^n$, $n=0,1,2...$, 
  and dashed contours at $-$3$\sigma$, where $\sigma$ is the r.m.s.\
  noise level (Table~1). The synthesised beam is shown in the bottom
  right corner.  In the {\it Planck} cosmology$^{30}$, 1\,arcsec
  corresponds to 8.4\,kpc at $z$ =2.3 (uncorrected for lensing magnification).  }
\end{figure}

\begin{figure}
\begin{center}
\includegraphics[angle=0,width=0.8\textwidth]{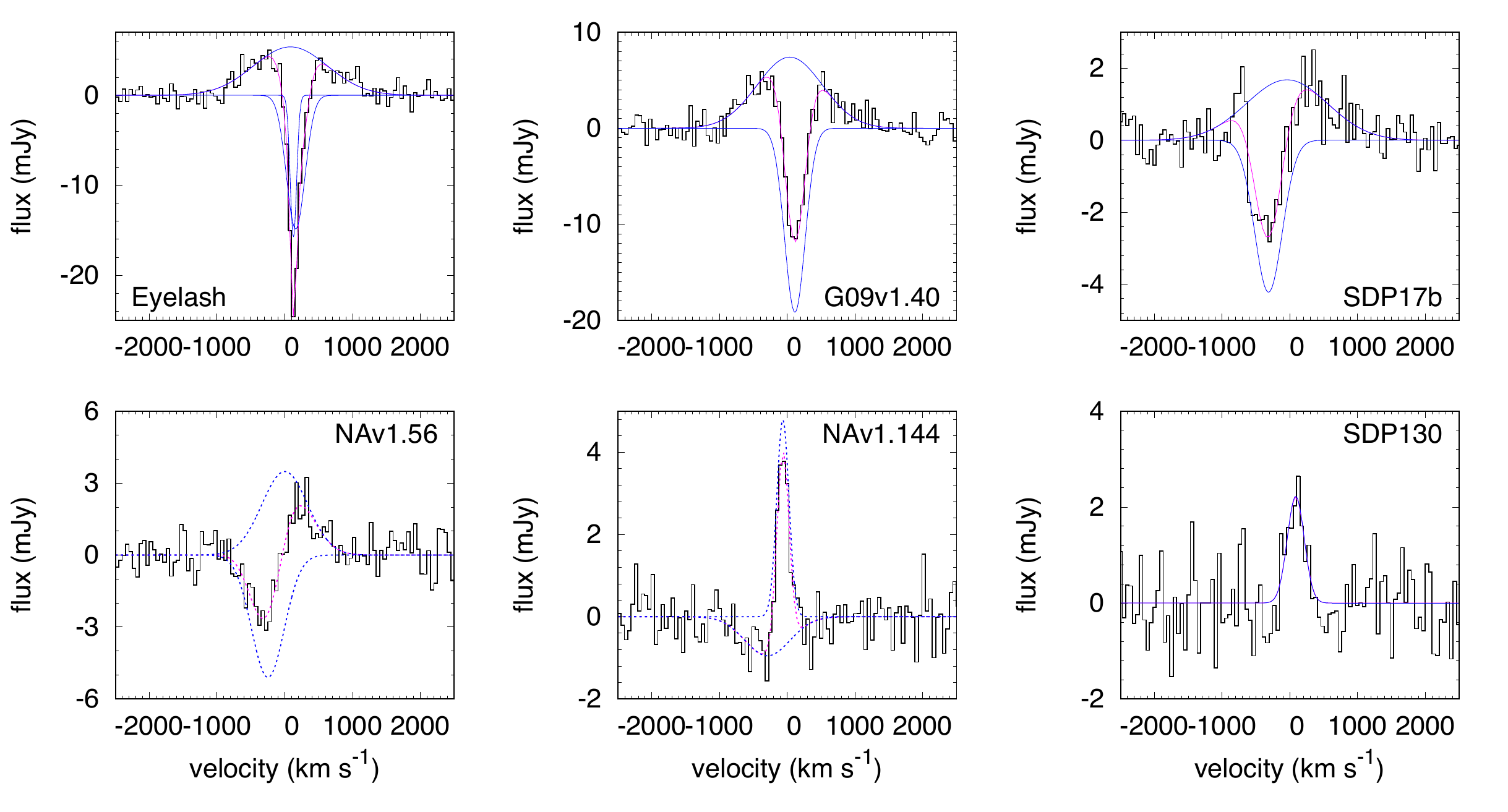}
\end{center}
\caption{{\bf ALMA \CHp\ $J$ = 1--0 spectra of our lensed targets. } 
  Continuum-subtracted \CHp\ $J$ = 1--0 spectra, integrated over the solid angle defined in the caption of Table~1.  
 The pink curves show the final fits;  the blue curves show the contributions of the emission (one Gaussian) and absorption (one or two Gaussians). The large velocity overlap of the absorption 
  and emission in NAv1.56 and NAv1.144 makes the fits only indicative (dotted lines). The velocity scale is with respect to the accurate redshifts of Extended Data Table~1. 
  The continuum  was removed with a linear baseline to perform the fits.}
\end{figure}

\newpage

\begin{table}
\small
\begin{center}
\caption[]{\bf{Properties of the lensed SMGs and ALMA \CHp\ line and continuum fluxes.}}
\bigskip
\smallskip
\begin{tabular}{llllllllll}
\hline \noalign {\smallskip}
Name$^a$  & $z$ $^a$  & $\mu$  $^b$         &$\log L_{\rm FIR}$$^c$    & $r_{\rm SMG}$ $^b$ & SFR$^d$                  & $S_{1.2\,{\rm mm}}$  &$\sigma_{1.2\,{\rm mm}}$   & $S_{\CHp}$$^e$  \\
                    &       &      &   \lsol                &  kpc         &  \msol yr$^{-1}$  &   mJy  & $\mu$Jy/beam &  mJy   \\
\hline \noalign {\smallskip}
Eyelash         & 2.3   & $\sim$35 $^{24}$               & 12.22 $\pm$0.07   &  $\sim$0.3$^f$                 & 285 &        38.15 & 30   &   5.4   \\
G09v1.40      & 2.0  &   15.3$\pm$3.5$^{16}$    & 12.37$\pm$0.09     & 0.41                       & 404  &       30.12 & 40  &    7.4   \\
SDP17b        & 2.3   &    6.2 $\pm$0.2               & 12.82$\pm$0.05      &  0.72                       & 1133  &       13.52 & 30  &     1.7   \\
NAv1.56        & 2.3   &    9.79$\pm$0.12            & 12.87$\pm$0.05       &   0.76                     & 1105   &        18.73 & 30   &     3.5   \\
NAv1.144      & 2.2   &  4.16 $\pm$0.12              & 12.89$\pm$0.08        & 0.84                       & 1338  &       18.57& 50   &   $>$   3.8  \\
SDP130        & 2.6   &  8.6$\pm$0.4                    & 12.33 $\pm$0.07     & 1.2                      & 367 &        7.4 & 40   &      2.  \\
\hline \noalign {\smallskip}
\end{tabular}
\end{center}
$^a$  The IAU full names that provide the galactic coordinates of the sources, as well as the full precision redshifts used are given in Extended Table 1. 
$^b$ Magnification and half-light size inferred from lens modelling of our ALMA data at rest-frame 360\,$\mu$m (see Methods), except for G09v1.40 and 
the Cosmic Eyelash where we adopt values from the literature. The uncertainty on the half-light sizes is $\sim$15\%.
$^c$ Intrinsic $L_{\rm FIR}$ integrated over 40--120\,$\mu$m$^{16}$, unlensed with our lens models. 
$^d$ Star formation rates assuming a Salpeter initial mass function$^{16}$  and no contribution from active galactic nuclei (AGN). These values are intended to illustrate only the differences among the galaxies.
$^e$ Observed peak \CHp\ line emission flux  integrated over a solid angle
defined by the contour level  0.05$\times$ the peak continuum level.
$^f$ Size in the source plane of  the brightest star-forming region (source X)$^{24}$.
\end{table}



\begin{table}
\small
\begin{center}
\caption[]{\bf{Characteristics of the \CHp\ lineprofiles and inferred properties of the  turbulent reservoirs}}
\bigskip
\smallskip
\begin{tabular}{llllllllll}
\hline \noalign {\smallskip}
Name & $v_{\rm em}$  $^a$ & $\Delta v_{\rm em}$ $^a$ & $\Delta v_{\rm abs}$ $^a$ & $v_{\rm off}$ $^a$  & $\tau_0$ $^a$ &  $N(\CHp)$ $^b$  & $r_{\rm TR}$ $^c$    & $M_{\rm TR}$ $^c$ & $L_{\rm TR}$ $^c$ \\
           & \kms  &  \kms               & \kms                   &   \kms      &               &  10$^{14}$\cq &  kpc    &  $10^{10}\msol$ &   $10^9\lsol$     \\
\hline \noalign {\smallskip}
Eyelash        & 30$\pm$53   & 1301$\pm$112  & 232$\pm$40    & 143$\pm$11             &1.2           &  8.3           &  8         & 1.4    &   1.2   \\
G09v1.40     & 28$\pm$34   & 1124$\pm$87    &  361$\pm$24    & 111 $\pm$7               & 0.70        & 7.4           &  12       &  1.1    &   2.3     \\
SDP17b      &  -21$\pm$96  & 1394$\pm$157    &  490$\pm$53  & -315$\pm$ 28            & 0.33      & 4.7            &  17       &  0.9   &    3.4      \\
NAv1.56       &       0$^d$            &  $>$ 814            & 579$^d$         & -249               & 0.29     & 4.9              & 20      &  0.8     &  4.5 \\
NAv1.144      &       0$^d$            &   $>$ 217          &    799$^d$    &    -227                       & 0.25         & 6.                 &   --    &        --     &      --                   \\
SDP130         & 85$\pm$21  &  248$\pm$88       & --                 & --                                & --              &                     &             &       &               \\
\hline \noalign {\smallskip}
\end{tabular}
\end{center}
$^a$ Parameters inferred from fits made simultaneously on the emission
and absorption lines: $\Delta v_{\rm em}$ and $\Delta v_{\rm abs}$ are
the FWHM,  $v_{\rm off}$ and $v_{em}$ are the offset velocity of the absorption Gaussian and the emission line centroid velocity
with respect to the systemic velocity of the galaxy as measured in CO,   
and  $\tau_0$ the absorption peak optical depth (see Methods).
 $^b$ Average  \CHp\ column density in a solid angle subtended by 
the background source of radius $r_{\rm SMG}$ (see Methods).
$^c$ Radius, mass and kinetic luminosity of the turbulent reservoirs  computed for $t_{\rm TR} = 50$\,Myr
(see Methods). 
$^d$ In these sources, the emission and absorption velocity coverages broadly overlap and the fits have been performed imposing $v_{\rm em}=0$. In NAv1.56, the convergence of the fit is poor. 
In NAv1.144, the strong residual emission is so narrow that the fit parameters are non-physical and have not been considered.
\end{table}

\clearpage

\newpage
\begin{addendum}

\item[Acknowledgements:] ALMA is a partnership of ESO (representing
  its member states), NSF (USA) and NINS (Japan), together with NRC
  (Canada) and NSC and ASIAA (Taiwan), in cooperation with the
  Republic of Chile. The Joint ALMA Observatory is operated by ESO,
  AUI/NRAO and NAOJ. NRAO is a facility of the NSF operated under
  cooperative agreement by Associated Universities, Inc. R.J.I.\
  acknowledges support from ERC in the form of the Advanced
  Investigator Programme, 321302, COSMICISM.
 EF and BG acknowledge support from 
  the national CNRS program {\it Physique et Chimie du Milieu Interstellaire} (PCMI).
 \item[Author contributions:] EF, EB, FB and DE conceived the initial scientific argument and wrote the
ALMA proposal with BG, MAZ, PMA, AO and RSB. MAZ reduced the ALMA data.
 BG and EF analyzed the spectra. BG provided the
results of the shock models. RJI, IOG and FW were invited to join the team at a later stage to provide the 
results of the lens models (IOG) and contribute to a year-long debate on the data interpretation. EF 
wrote the paper with contributions from all the authors.
   \item[ Author informations:] The authors declare that they have no
competing financial interests. Correspondence and requests for materials
should be addressed to E.F. (email: edith.falgarone@ens.fr).

\end{addendum}

\newpage

\section*{Methods}

\subsection{1 - ALMA observations}
The six targets were observed with ALMA in Cycle 2 as part of project
2013.1.00164.S. The \CHp(1-0) line ($\nu_{\rm rest}$ = 835.08\,GHz) 
at redshifts $z \sim$ 2.0--2.6 is shifted into ALMA band 6 at
frequencies between 230 and 270 GHz. The ALMA baseline correlator was configured to
use three basebands in the upper sideband, so as to cover a maximum contiguous
velocity range of approximately 5000 \kms\ centred on the \CHp\ line. The low
resolution Time Division Mode was used, which, using spectral averaging, provides 240 channels of each 7.812 MHz 
per baseband. This corresponds to a spectral
resolution of 19 \kms\ after online Hanning smoothing. Between 32 and 34 antennas were used for these
observations. A standard observing strategy was used, where a phase reference source
was observed every 8 minutes. 
The total on source time for the six sources combined was 195 minutes.
Absolute flux calibration was obtained through the
observation of solar system objects and quasars with well-known flux values. 
The data calibration followed standard routines and was carried out using the CASA 4.4
(common astronomy software applications) package. Continuum images were made for
each field using a 'robust' weighting scheme with a Briggs parameter of 0.5. All
targets were sufficiently bright to perform one round of phase-only and one round of
amplitude and phase self-calibration, which significantly improved the image quality
and noise level. These self-calibrated data sets were subsequently used to construct
data cubes centred around the redshifted \CHp\ frequency, with a channel spacing of 50
\kms. The typical spatial resolution of the final data cubes is 0.5 arcsec, and the
r.m.s. noise level is in the range 0.25 to 0.4 5 mJy per channel, depending on the data cube. \\
This paper makes use of the following ALMA data: ADS/JAO.ALMA\#2013.1.00164.S.

\subsection{2 - Targets}

{\bf The Eyelash:} This three-image source strongly lensed by a foreground cluster has a combined amplification of 37.5$^{24}$.
High-resolution interferometry with the Sub-Millimeter Array (SMA) resolved the rest-frame 260 $\mu$m continuum emission in two
sets of four clumps. The clumps have physical (source-plane) sizes of 100--300\,pc 
in a region $\sim$ 2\,kpc within the background galaxy. 
The total gas mass inferred from CO is  $M_{\rm gas}\sim 4 \times 10^{10}$\msol, the stellar mass is $M_\star= 3 \times 10^{10}$\msol$^{24}$, 
giving a gas mass fraction $f_{\rm gas}=M_{\rm gas}/(M_\star+M_{\rm gas}) \sim 60$\%. 
The CO spectral line energy distribution suggests that this is a multiple component system
with different conditions$^{31}$. Low density gas at low temperature is required to model the low-lying transitions of CO. The OH absorption line$^{20}$   is 
interpreted as an outflow originating in the largest star-forming region of the galaxy.

{\bf The  H-ATLAS sources:} 
The redshift determination in G09v1.40 from the \wat\ lines$^{32}$, $z$=2.093, is different from the centred value$^{16}$ of the  \CHp\ emission line. 
880 $\mu$m  SMA observation of the dust continuum shows two close components with a separation of $\sim 1$ arcsec 
along the east and west direction. The lensing model$^{16}$ gives $\mu = 15.3 \pm 3.5$. Both the {\it 10m-Keck} telescope and the {\it Spitzer} space telescope images, in the near-IR and mid-IR respectively, agree about
a stellar component magnification of $\mu$ = 11.4+0.9/-1$^{33}$.
For all the others, the results of the lensing model applied to our ALMA data (Table~1) differ by $\sim 20$\% from those inferred from the SMA data$^{16}$, 
except for SDP130 where the ALMA-inferred magnification is $\sim 4 \times $ that from SMA.
NAv1.56 is a possible merger at a projected separation of 29 kpc (3.5 arcsec).
 In SDP17, the total gas mass inferred from CO is $M_{\rm gas} = 5.9 \times 10^{10}$\msol, the stellar mass is $M_\star= 2.4 \times 10^{11}$\msol$^{25}$,
so that the gas mass fraction $f_{\rm gas} \sim 20$\%. \\
The IAU names of the targets, redshifts and CO and \wat\ lines characteristics are given in Extended Data Table~1.
 
\subsection{3 - Exploitation of  the \CHp\ lineprofiles}

\CHp\ has such a highly endoenergic formation route via C$^+$ + H$_2$ ($E_{\rm form} \sim$ 0.4\,eV) 
that a specific chemistry activated by supra-thermal processes is mandatory$^{7-9}$. 
Its fast destruction due to collisions with H and \HH\  makes its lifetime extremely
short, 
even in dilute gas.
Being a light hydride with high dipole moment,
the critical density of its $J$ = 1--0 transition is $\sim 10^7$\,\cc,
about 10$^4\times$ larger than that of CO $J$ = 1--0, so 
in low-density  gas, most molecules are in the ground state providing large $J$ =1--0 line opacities.  
In absorption, \CHp\ is therefore a
tracer of dissipation of mechanical energy in low-density gas ($n_{\rm H}<10^2$\cc).
In contrast, in emission, \CHp\ $J$= 1--0 line is a tracer of 
dense ($n_{\rm H}>10^5$\cc) ultaviolet(UV)-illuminated gas$^{12}$, such as UV-irradiated shocks because, unlike most molecules, the
abundance of \CHp\ is enhanced in intense UV fields$^{10,11}$. \\
All our targets are lensed starburst galaxies, with most likely different magnification for the dust 
continuum emission, the gas seen in absorption and that seen in emission.
The angular and spectral resolution of our ALMA data allow some information on the respective distributions of 
the absorption and the emission in velocity (Extended Data Fig. 1) and of the continuum dust emission and \CHp\ absorption 
and emission in space (Extended Data Figs. 2 and 3). To interpret these data, modelling of differential lensing is warranted. 
However, the optical depth estimates of the absorption lines  are not affected by differential lensing because 
the background dust emission and the absorbed emission experience the same lensing.
Fits have been performed simultaneously on the emission
and absorption lines, and 
the uncertainties on the parameters inferred (Table~2) have been evaluated with the bootstrap method to take
into account the correlation between these parameters, in addition to the data noise level.  
The broad  overlap between the absorption and emission velocity ranges makes  the uncertainty on 
the absorption linewidth larger than 100\% for NAv1.56.  For NAv1.144, no solution with broad
emission is found by the fitting procedure, although it is physically plausible. \\
The emission line luminosities, computed as 
$$
L_{\CHp}=1.04 \times 10^{-3} S_{\CHp} \Delta v_{\rm em} \nu_{\rm rest} D_{\rm L}^2/ (1+z)
$$ 
where $D_{\rm L}$ is the luminosity distance expressed in Mpc (Extended Data Table~1), $S_{\CHp}
\Delta v_{\rm em}$ is  the line-integrated flux in Jy\kms\ and $\nu_{\rm rest}$ = 835.07895\,GHz, are in the range 2--7$\times 10^8 \lsol$.
The  absorption line optical depth, computed at the velocity $v_0$ of the absorption peak, is derived from these fits:
  $$
  \tau_0=-\ln(S_{\rm abs}(v_0)/[S_{\rm cont}+S_{\rm em}(v_0)],
  $$
   where $S_{\rm cont}$, $S_{\rm abs}$ and $S_{\rm em}$ are the continuum, absorption and emission fluxes respectively. \\
The \CHp\ column densities are the 
average  values in a solid angle subtended by the area, $\pi
r_{\rm SMG}^2$, of the background source assumed to be uniformly covered by the absorbing screen:
$$
N(\CHp)=3\times10^{12}\cq\ \tau_0 \,(\Delta v_{\rm abs}/1\kms) 
$$
where $\Delta v_{\rm abs}$ is the FWHM of the absorption line.\\
The formal errors on $N(\CHp)$ essentially follow the
signal-to-noise ratio of the data.
The velocity resolution is coarse and we cannot rule out the
possibility that portions of the line might be saturated. The estimated values of $N(\CHp)$
could  therefore be lower limits, so in Table~2 we provide them without a compiled
error.

\subsection{4 - The turbulent framework }

Since its discovery in the diffuse interstellar medium (ISM)$^{34}$, the  puzzle raised by the large  \CHp\ 
abundances had focused on our Solar Neighbourhood because the line used was an electronic transition  
($\lambda \sim 0.4 \mu$m) detected in absorption against nearby stars.
The rest frequency of \CHp($J=$1-0) falling in a domain of high atmospheric opacity, the first observations of \CHp(1-0) were achieved with the  {\it Herschel} satellite. The 
 line was detected in absorption against the dust continuum of remote Galactic star-forming regions with {\it Herschel}/HIFI $^{11,35,36}$ and in absorption or emission with {\it Herschel}/SPIRE$^{37}$. 
 The {\it Herschel}/HIFI data reveal the large opacity of the \CHp(1-0) line in absorption in the Milky Way and show that the \CHp\ absorption profiles differ considerably from those of other lines$^{36}$. 
The  {\it Herschel}/SPIRE detections in nearby starbursts$^{38-40}$ lack spectral resolution to definitely disentangle emission and absorption but Arp220 has a strong absorption line while Mrk231 
has the line in emission.
A recent detection of the visible lines of \CHp\ against a supernova in M82 reveals a highly turbulent component$^{41}$. \\
In the Milky Way, the \CHp\ abundances observed in the diffuse molecular gas were found to scale with the turbulent energy transfer rate per unit volume, $\epsilon$.
The models of turbulent dissipation regions provide an expression of this scaling$^{9}$. 
In a turbulent cascade, the dissipation rate per unit volume is on average equal to the energy transfer rate per unit volume in the cascade: $\epsilon_l= \frac{1}{2} \rho_l {\overline v_l}^2/ \tau_l$ where $\tau_l=l/{\overline v}_l$ is the dynamical time of scale $l$, and 
${\overline v}_l$ and $\rho_l$ are the mean velocity and  mean mass density at that scale, respectively. It is this quantity, 
$$
\epsilon_l=\frac{1}{2} \rho_l {\overline v_l}^3/l,
$$
 that is observed to be uniform on average, albeit with a huge scatter, in the inner Galaxy$^{42}$ and is found to be an invariant in numerical simulations of compressible turbulence$^{43}$. This quantity is
 shortened into "turbulent energy flux" in what follows because it is a flux in the wavenumber-space of the turbulent cascade. \\
The fraction of the turbulent energy flux required to sustain the observed number of \CHp\ molecules in the volume $\pi r_{\rm bg}^2 L$ subtended by the 
background continuum source (radius $r_{\rm bg}$) and the length of the line of sight $L$ is $\alpha=\dot{E}_{\CHp}/\dot{E}_{\rm turb}$, with
$$
\dot{E}_{\CHp}= N(\CHp) \pi r_{\rm bg}^2 E_{\rm form}/t
$$ 
and 
$$
\dot{E}_{\rm turb}= {\overline \epsilon} \times  \pi r_{\rm bg}^2 L. 
$$
The turbulent energy flux $\overline \epsilon$ is a large scale 
 average and  the \CHp\ lifetime is $t=1$yr $f_{\HH}^{-1} (n_{\rm H}/50\cc)^{-1}$, where $f_{\HH}=2n(\HH)/n_{\rm H}$ is the \HH\ fraction and $n_{\rm H}$ and $n(\HH)$ are the proton and \HH\ densities, respectively.
This is why  the \CHp\ opacity of absorption lines that provides $N(\CHp)$ is a direct measure of the turbulent energy flux in the medium. 
Introducing $N_{\rm H,abs}={\overline n}_{\rm H} L$,  the total  column density of gas of average density ${\overline n_{\rm H}}$ sampled by the absorption, the observed abundance 
is $X(\CHp)=N(\CHp)/N_{\rm H,abs}$, and  therefore $\alpha= X(\CHp) \times {\overline n_{\rm H}}/{\overline \epsilon} \times  E_{\rm form}/t$.
In the Solar Neighbourhood$^{9}$, the average values are  $\langle X(\CHp) \rangle =7.6 \times 10^{-9}$, ${\overline n}_{\rm H} \sim 1$\cc, and $\overline \epsilon \sim 10^{-25} \eccs$, so that $\alpha= 7.7 \times 10^{-4} f_{\HH} n_{50}$, where $n_{50}=n_{\rm H}/(50\cc)$  is the local gas density.
Galactic observations of [CII] absorption and CI emission along the same background sources where \CHp\ 
absorption is observed show that this gas is indeed the cold neutral medium (CNM)$^{44}$. The average \HH\ fraction in this low-density molecular gas is found to be $\overline{f}_{\HH} = 0.4$ from CH and HF
absorption lines along the same sightlines$^{36}$, with a density in the range $n_{\rm H} =$ 30--50\,\cc. The filling factor of the CNM in which \CHp\ forms,
$f_v={\overline n}_{\rm H}/n_{\rm H}$,  is much less than unity.

\subsection{5 - Determination of the turbulent reservoir  properties}

 We follow the same line of argument to study the turbulent gas around the SMGs. 
 The turbulent approach  is supported by the fact that the narrowest [broadest] absorption components have the smallest [largest] velocity offsets ($v_{\rm off}$, Table 2).
  We also rely  on a fundamental property of turbulence -- a pencil beam across a turbulent volume samples all the velocities because of the velocity correlations --
 to relate the size of the turbulent reservoir  to the mean turbulent velocity provided by the absorption linewidth. 
 We assume a negligible contribution of rotation to the absorption linewidths (see Extended Data Fig. 1).
 The unknown \CHp\ abundance is therefore an average value over the volume sampled by the absorption, in the same way as 
 in the  Milky Way. \\
We again express that  a  fraction $\alpha$ of the turbulent energy flux 
$$
\dot{E}_{\rm turb}={1 \over 2} M_{\rm abs} \overline{v}_{\rm abs}^3/  r_{TR}
$$
 in the gas mass sampled  by the absorption, $M_{\rm abs}=\mu_{\rm p} m_{H} N_{\rm H,abs} \pi r_{\rm SMG}^2$,  
 is sufficient to sustain the number of \CHp\ molecules observed in absorption against the background continuum source of radius $r_{\rm SMG}$,    
  $$
  \dot{E}_{\CHp}=  N(\CHp)\, \pi r_{\rm SMG}^2\, E_{\rm form}/t.
  $$
  We thus obtain the relation between the radius of the turbulent reservoir and the \CHp\ abundance:
  $$
  r_{\rm TR} \, X(\CHp) =\alpha \frac{\mu_{\rm p} m_{\rm H}}{2} \frac{t}{E_{\rm form}} {\overline v}_{\rm abs}^3
  $$ 
where  the mean turbulent velocity is ${\overline v}_{\rm abs}=0.7 \Delta  v_{\rm abs}$ and $\mu_{\rm p}$ is the mean mass per particle. 
 \CHp\ abundances in the range (0.8--5.2)$\times 10^{-8}$ follow from the estimated values of $r_{\rm TR}$,  and hydrogen column densities in the range 
10$^{23}$--9$\times 10^{21}$\,\cq\ are inferred.
The mass of the turbulent reservoir is then computed by assuming mass conservation across the different shells, regardless of the global inward/outward 
or turbulent motions, so that the average density of the reservoir decreases with radius as $\overline{n}_{\rm H} \propto r^{-2}$. 
 As a result, the hydrogen column density at impact parameter $b$ decreases as $N_{\rm H}(b) \propto b^{-1}$ and the mass is 
 $M_{\rm TR}=2\pi\ \mu_{\rm p} m_{\rm H} r_{\rm TR} r_{\rm SMG} N_{\rm H,abs}$. \\
 Introducing $\Delta v_{400}=\Delta  v_{\rm abs}/(400\kms)$ and $N_{14}=N(\CHp)/(10^{14}$\cq) for the observed quantities, so that $N_{14}/\Delta v_{400}=12\tau_0$, $X_8=X(\CHp)/10^{-8}$ for the inferred abundances, 
 $\alpha_{3}=\alpha/10^{-3}$ and $t_{50}=t_{\rm TR}/(50$\,Myr) for the most probable age of the turbulent reservoirs, with $t_{\rm TR} \sim  t_{\rm SMG}/2 \sim 50$\,Myr, the radius and mass of the turbulent reservoirs are
 $$
 r_{\rm TR} = 14.0 \,{\rm kpc}\, \times  \Delta v_{400}\, t_{50}
 $$ 
 and 
 $$
 M_{\rm TR} = 3.8 \times 10^{10} \msol\ \times \tau_0  \, t_{50}^2 \,  r_{\rm SMG,kpc} (f_\HH\ n_{50}/\alpha_3) 
  $$ 
 and the \CHp\  abundance is 
 $$
 X_8 = 2.5 \, \Delta v_{400}^2 \, \alpha_{3} /(t_{50}\, f_\HH\  n_{50}).
 $$
  A mean mass per particle $\mu_{\rm p}=1.27$ is adopted. 
 These expressions help distinguish the dependence of the inferred sizes and masses on the observed quantities 
 ($\Delta  v_{\rm abs}$ and $N_{14}$) and free parameters ($f_\HH$,  $n_{50}$ and $\alpha_3$).
 The local gas density $n_{50}$ and the \HH\ fraction 
introduce some degeneracy. However, the range of possible values is limited. From the Galactic studies, $\overline{f}_{\HH}=0.4$ and the gas densities
of the \CHp-rich regions are in the range 30 -- 50 \cc. All of the determinations are therefore within factors of a few.
The \CHp\ abundances are close to the Galactic values, as would be expected if the \CHp\ production rate were comparable to that in the Galaxy 
and the metallicities were simlar to solar values, as is observed$^{45}$  in star-forming galaxies at $z\sim$ 2.\\
 The kinetic luminosity is equal to the energy input rate needed to balance the dissipation of the  kinetic  energy 
 $E_{\rm turb}=\frac{1}{2} M_{\rm TR} {\overline v}_{\rm abs}^2$ 
 in a dynamical time $t_{\rm TR}=r_{\rm TR}/{\overline v}_{\rm abs}$ or, 
 with the appropriate units, 
 $$
 L_{\rm TR}=2.6\times 10^{9} \, \lsol\ \,M_{\rm TR,10} \, \Delta v_{400}^2/ t_{\rm 50},
 $$
  where $M_{\rm TR,10}=M_{\rm TR}/(10^{10}$\msol).  
The mass outflow rate at velocity $v_{\rm out}$ that is required to balance turbulent dissipation,
 $L_{\rm TR}= \frac{1}{2} \eta \dot{M}_{\rm out} v_{\rm out}^2$ (assuming an efficiency $\eta < 1$ of the energy transfer) is 
  $$
  \dot{M}_{\rm out} \sim \frac{19}{\eta} \msol\ {\rm yr}^{-1}\, L_{9} \,/v_{800}^2
  $$ 
  where  $L_{9}=L_{\rm TR}/(10^9 \lsol)$ and  $v_{800}=v_{\rm out}/(800\kms)$. \\
 The turbulent reservoirs are more than 10 times larger  the SMGs. Such kpc-scale haloes of low-density gas, comprising a neutral  diffuse molecular phase, 
 have been invoked to interpret the low-lying OH absorption  lines in Arp\ 220$^{46}$ and NaI D absorption lines$^{47}$ in ULIRGs, as well as CO(1-0) lines broader than CO(3-2) lines 
 in distant SMGs$^{48}$. Large kpc-scale haloes of atomic hydrogen are being discovered in the Ly$\alpha$ line around high-redshift galaxies$^{29,49}$.
 Last, the gas of the turbulent reservoirs is multi-phasic because its average density $\overline{n}_{\rm TR} = N_{{\rm H,abs}}/r_{\rm TR}\sim$ 0.2-- 6\cc, is 
 less than that of the \CHp-rich gas at $n_{\rm H} \sim $30--50\cc,  that fills only a low ($<10$\%) fraction of the volume.

\subsection{6 - Models of irradiated shocks}

A medium shocked at high velocity emits UV photons that interact with the pre-shock gas$^{26}$. 
Depending on the shock velocity, this radiative precursor heats, 
dissociates (for $v_{\rm sh} >$ 90 \kms) or even ionizes (for $v_{\rm sh} >$ 120 \kms) the upstream medium, hence 
preventing the formation of molecular species. Therefore, while \CHp\ emission has very broad line 
profiles, it is necessarily formed in shocks of much lower velocity. 
The extremely broad \CHp\ emission lines trace the large velocity dispersion of lower-velocity shocks, as observed in the kiloparsec-scale shock of the Stephan's Quintet$^{27,50}$.\\
The production of \CHp\  in low-velocity interstellar shocks has been studied using the 
Paris-Durham shock code, a state-of-the-art model that was recently improved to take into account 
the basic physics of photo-dominated regions$^{8}$. 
In these starbursts,  UV-illumination comes from star formation with a possible contribution from  
the thermal emission of the optically thick accretion disks of AGNs. 
Using observations of 22 AGNs$^{51}$, we estimate that the luminosity of a typical AGN 
with redshift $z>1$ is comparable to that due to high-mass star formation, taking dilution within the starburst into account. 
We ran the code for a range of values of 
the preshock density (10$^3$ -- $10^6$ cm$^{-3}$), the UV radiation field (1 -- 
$10^4 \times$ the solar neighbourhood value), and the shock velocity (10 -- 80 
\kms), assuming a magnetic field of 2$[n_{\rm H}/1\cc]^{1/2}\mu$G, and a visible 
extinction (i.e. shielding from UV photons) of 1 magnitude.\\
We find that the critical velocity, above which stationary $C$-type shocks 
cannot propagate, strongly depends on the density and the radiation field. Below a 
critical value ($v_{\rm sh}\sim40$\,\kms\ i.e. $C$-type shocks), the large abundance of \HH\ and the 
ion-neutral decoupling are sufficient to produce column densities of \CHp\  as large as about
$10^{14}$ cm$^{-2}$ in the shock front, within timescales smaller than the expansion 
timescale of the gas. Above this critical value (i.e. $J$-type shocks), the rapid and 
self-sustained collisional dissociation of \HH\  prevents the immediate formation of \CHp\
in the post-shock where \HH\ reforms and the temperature is still high 
enough. Column densities of \CHp\  as large as those of $C$-type shocks are obtained for a 
kinetic energy input 5 $\times$ larger, over distances 100 $\times$ larger and 
for timescales larger than the expansion timescale of the post-shock gas.
We conclude that  $J$-type shocks are 
far less efficient than $C$-type shocks to produce \CHp.\\
Large \CHp\ column densities, up to  $N(\CHp) \sim 10^{14}$\cq\ and spread over several \kms\ per shock, are formed in $v_{\rm sh}=20$\kms\ {\it C}-type shocks 
propagating in dense pre-shock gas ($n_{\rm H} \sim 10^4$\cc). 
Lines as bright as 1\,mJy/beam, estimated in the Large Velocity Gradient approximation$^{52}$,  are obtained over a broad range of shock densities $\geq10^5$\cc\ and gas temperatures $\geq 100$K. This 
result does not greatly depend on the gas density. The thickness of each shock is very small, about a few hundred astronomical units. The number of shocks in the beam is difficult to estimate 
because it depends on the actual geometry of the shock ensemble and its distribution in velocity. 
Corrected for lensing with the magnifications estimated for the continuum emission, the intrinsic \CHp\  peak brightnesses are about 0.1\,mJy/beam.  Hence, millions of low-velocity shocks are required in the beam to reproduce the observed \CHp(1-0) brightness
spread smoothly over $\sim$ 1,000\kms. The \CHp(1-0) line opacity of individual shocks is less than unity; therefore there is no limitation on the emitted line intensity due to radiative transfer. \\
The lack of detection of CO(1-0) lines as broad as the \CHp(1-0) lines in those of our targets (except NAv1.56) where CO(1-0) has been observed (Extended Data Table 1) is due to the much larger emissivity of the \CHp(1-0)  transition compared to that of CO(1-0). The Einstein coefficients  for spontaneous emission are $A_{10}= 6.4 \times 10^{-3}$ s$^{-1}$  for \CHp(1-0)$^{12}$ and $A_{10}= 7.4 \times 10^{-8}$ s$^{-1}$ for CO(1-0). For an optically thin line, an upper limit of the $i\rightarrow j$ line integrated intensity 
is provided by $N_i \, A_{ij} h\nu_{ij}$, so the CO column density in the $J=1$ level has to offset the huge ratio $A_{10} \nu_{10} (\CHp)/ A_{10} \nu_{10} ({\rm CO}) \sim 1.6 \times 10^6$. 
Therefore as long as the column density of CO in the shocks does not exceed that of \CHp\ by a factor of about $10^7$, which is the case in the low-velocity $C$-type shocks, the CO(1-0) line remains very weak
and probably below the detection level.


{\bf Data Availability Statement:} This paper makes use of the following ALMA data:\\
ADS/JAO.ALMA\#2013.1.00164.S. 
The datasets generated and/or analysed during the
current study are available in the ALMA archive
(http://almascience.eso.org/aq/?project\_code=2013.1.00164.S) and are also available
from the corresponding author upon reasonable request.

\newpage

\setcounter{table}{0} 
\renewcommand{\tablename}{Extended Data Table }

\begin{table}
\small
\begin{center}
\caption[]{\em{Additional properties of the lensed SMGs}}
\begin{tabular}{cccccccc}
\hline \noalign {\smallskip}
Name & IAU name & $z$ &  $D_{\rm L}$ $^a$  & CO line &      $S_{\rm CO}$ $^c$        & $\Delta v_{\rm CO}$     &  $\Delta v_{\rm \wat}$   \\
                    &       &      &    Gpc &   &    Jy \kms\         &  \kms\         &   \kms\       \\
\hline \noalign {\smallskip}
Cosmic Eyelash  &   SMMJ2135-0102            &  2.3259   & 18.95     & 1-0   &  2.16$\pm$0.11 &       290$\pm$ 30     &        \\
G09v1.40           & J085358.9+015537$^b$    & 2.0894    &  16.97    &  4-3  &  7.5$\pm$2.3      &  198 $\pm$ 51    &    277$\pm$14 \\
SDP17b            & J090302.9-014127  $^b$     & 2.3051   & 18.95     &   1-0 &   0.9$\pm$0.1  &    320$\pm$ 10        &  250$\pm$60 $^{53}$   \\
NAv1.56            &  J134429.4+303036  $^b$     & 2.3010 &  18.95     &  1-0 &    1.1$\pm$0.1   & 1140$\pm$130           & 593$\pm$56 $^{32}$    \\
NAv1.144          &  J133649.9+291801 $^b$    & 2.2024   &  17.95   &  1-0 &    0.9$\pm$0.1   & 220$\pm$20            &  200$\pm$50 $^{32}$ \\
SDP130            &   J091305.0-005343  $^b$     & 2.6256  &  21.98    &  1-0  &   0.71$\pm$0.07 & 360$\pm$40          &     \\
\hline \noalign {\smallskip}
\end{tabular}
\end{center}
$^a$ Luminosity distances computed for $H_0=67.3$\kms\,Mpc$^{-1}$,
  $\Omega_{\rm M}=0.315$ and $\Omega_\Lambda=0.685^{30}$. \\
$^b$  {\it H}-ATLAS sources$^{12}$, redshift determinations$^{14}$  \\
$^c$  The CO(1-0) results on the {\it H}-ATLAS sources are from GBT/Zspec observations$^{21}$\\

\end{table}

\setcounter{figure}{0} 
\renewcommand{\figurename}{Extended Data Figure}

\begin{figure}
\includegraphics[angle=0,width=0.8\textwidth]{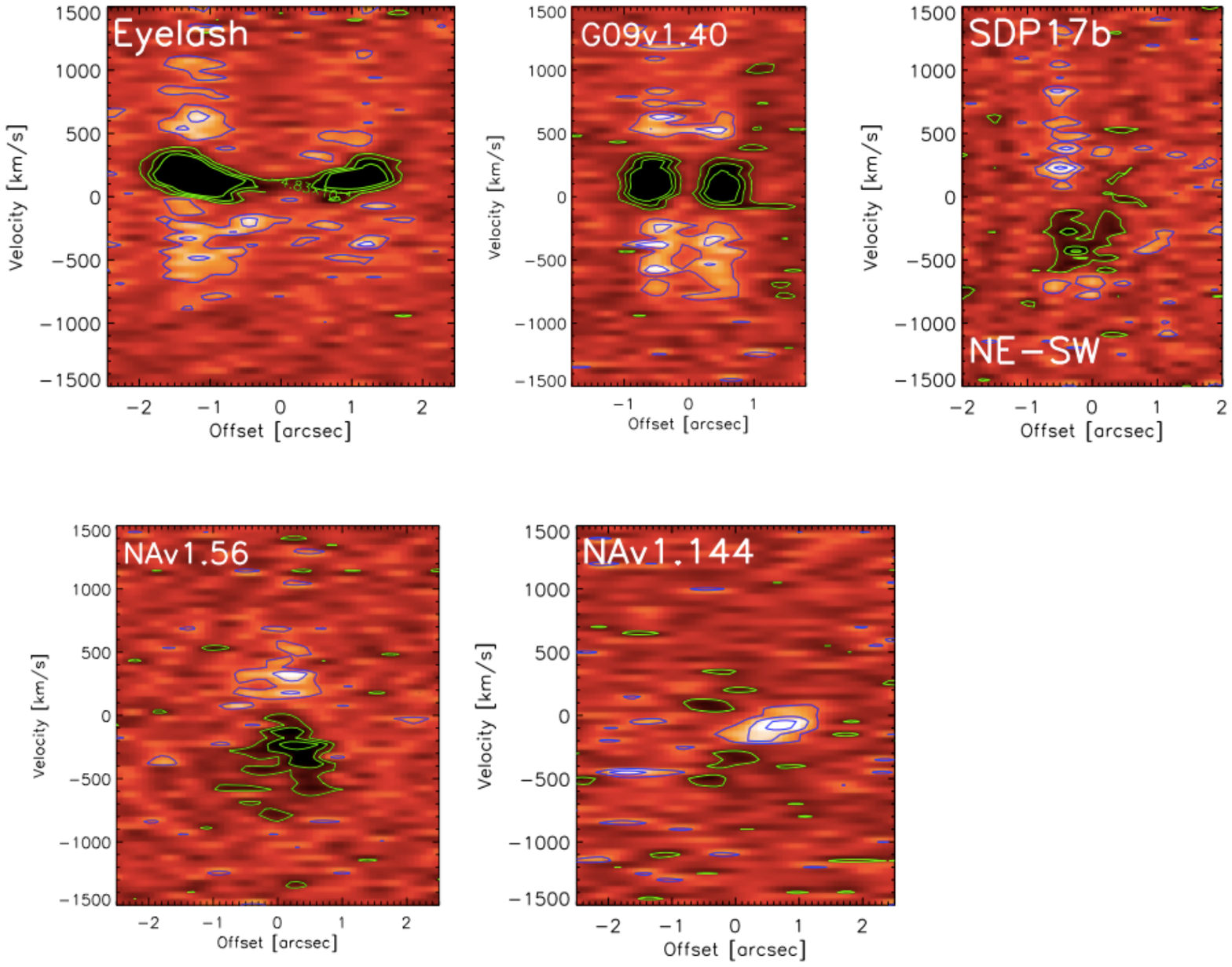}
\caption{{\bf Position-velocity diagrams of \CHp\ emission and absorption along selected cuts across the
  sources.}   The cuts are made along East-West directions, except for the Eyelash
  where it is made along the long-axis of the lensed images and SDP17b where it is along a NE-SW direction.
  \CHp\ emission appears in white (blue contours) and absorption in black (green contours). The first contour level and steps
 are  2$\sigma$. A velocity gradient is seen in the Eyelash absorption, twice smaller than that detected in CO$^{24}$.  }
\end{figure}

\begin{figure}
\includegraphics[angle=0,width=0.7\textwidth]{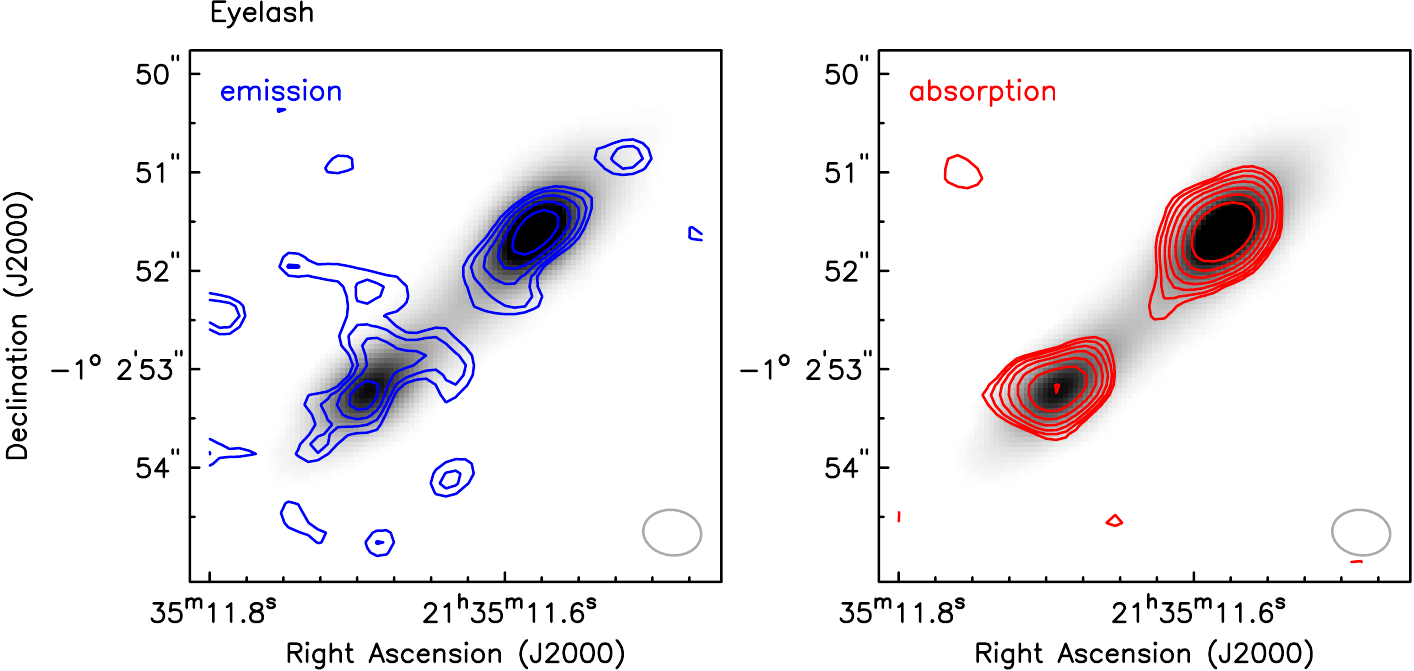}
\includegraphics[angle=0,width=0.7\textwidth]{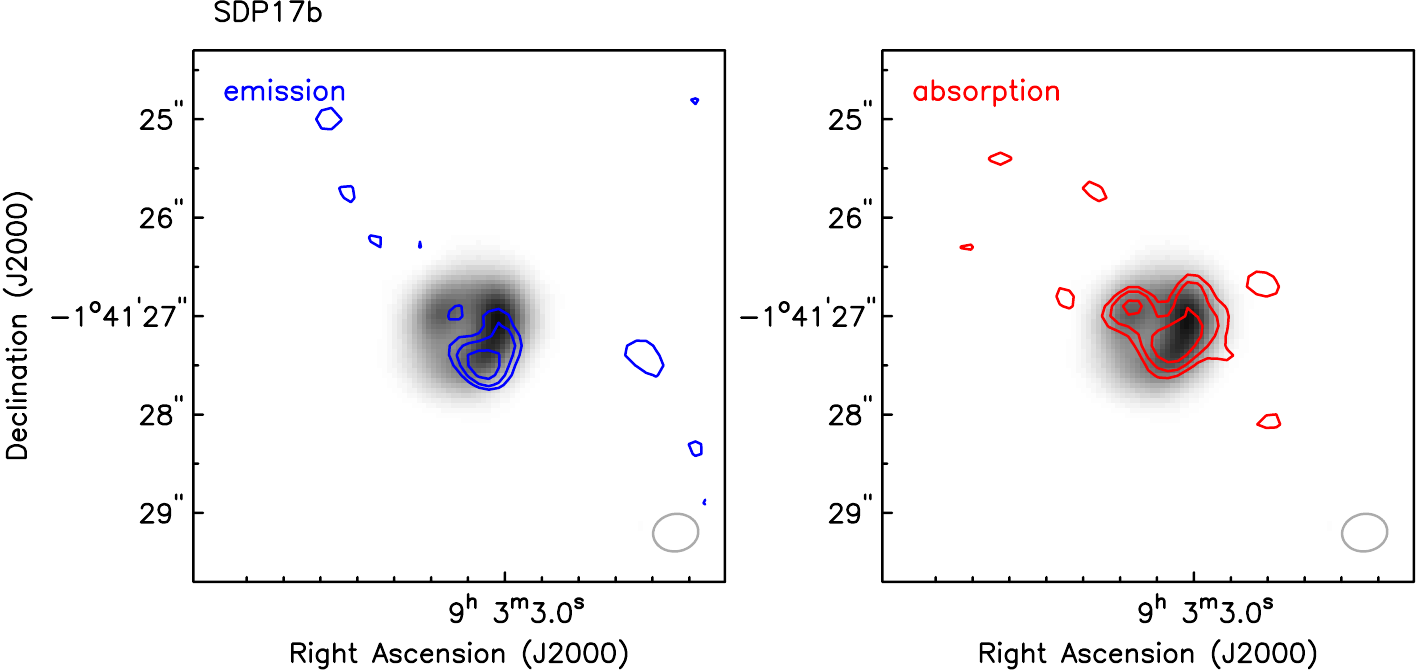}
\includegraphics[angle=0,width=0.7\textwidth]{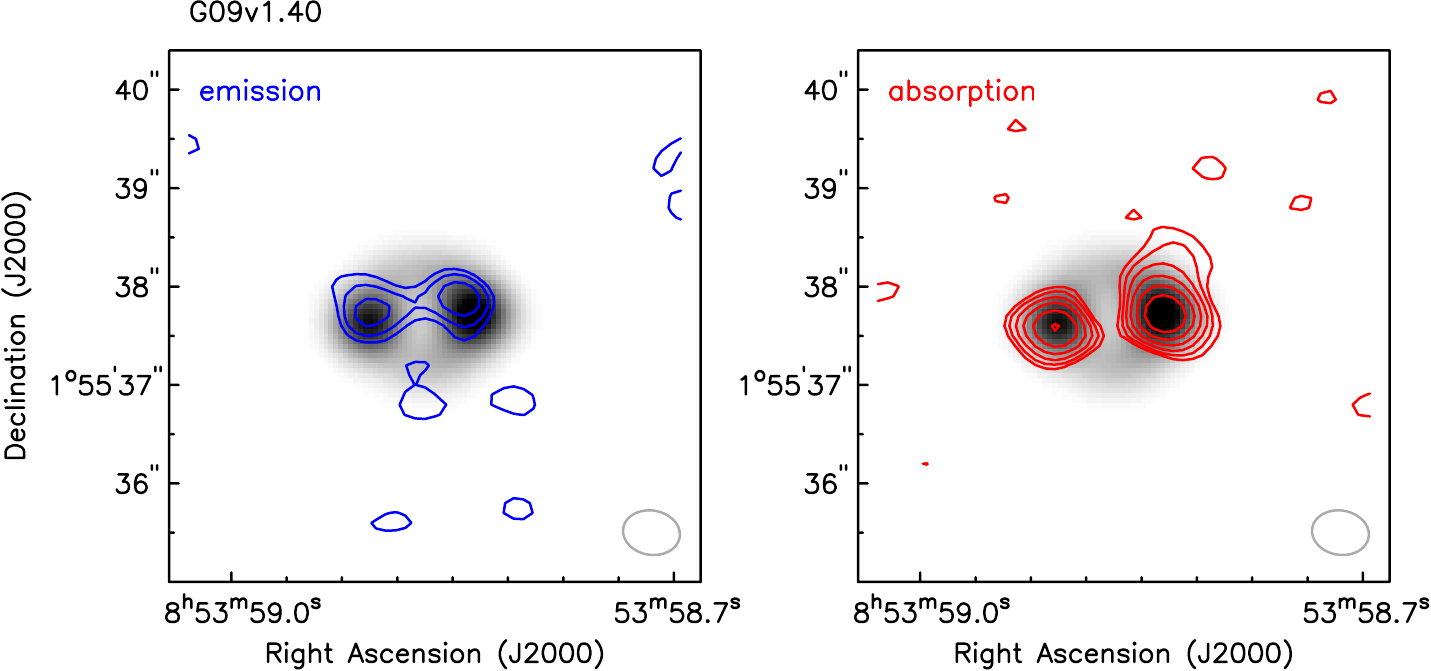}
\caption{{\bf \CHp\ emission and absorption overlaid on dust continuum emission for the Eyelash, SDP17b and G09v1.40.} 
The \CHp\ 
line-integrated emission (blue contours) and absorption (red
  contours), with contour levels in steps of 2$\sigma$, is overlaid on continuum
  emission (grey scale).  All of the images are lensed, so the differences between the distribution of 
  dust continuum and \CHp\  line emission are affected by differential lensing.  } 
\end{figure}

\begin{figure}
\includegraphics[angle=0,width=0.7\textwidth]{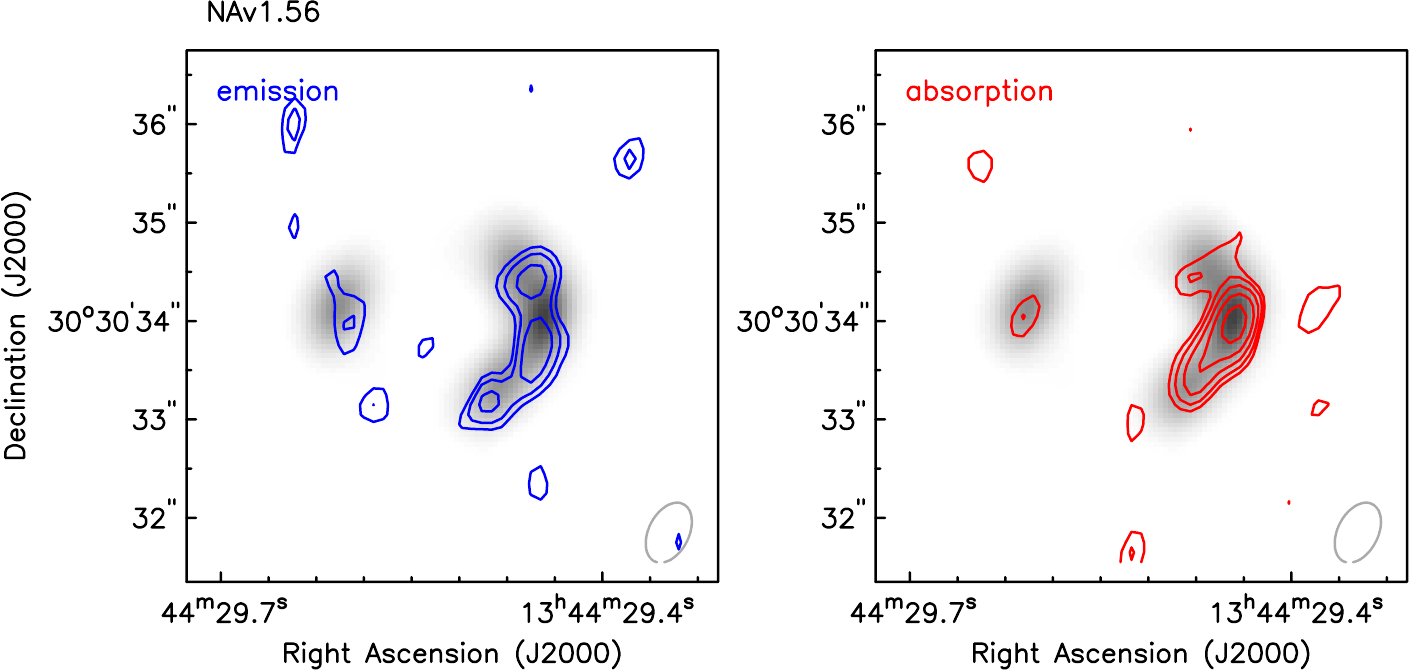}
\includegraphics[angle=0,width=0.7\textwidth]{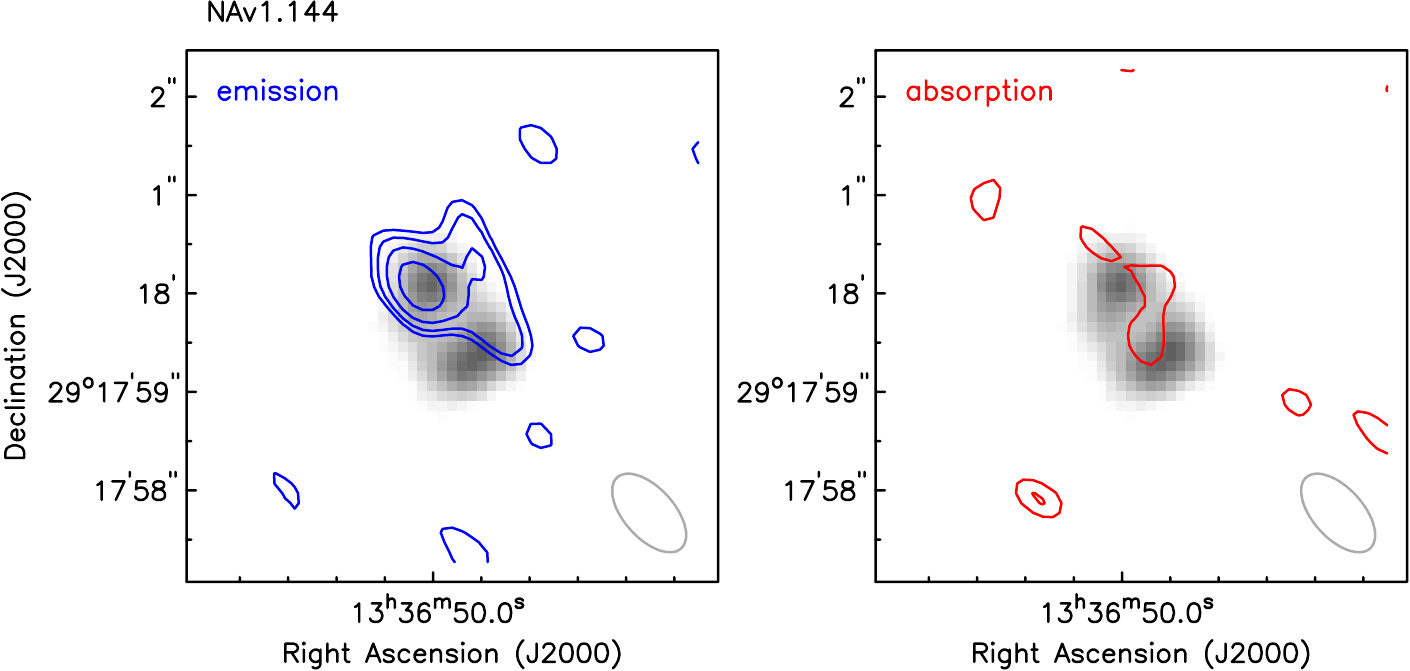}
\includegraphics[angle=0,width=0.35\textwidth]{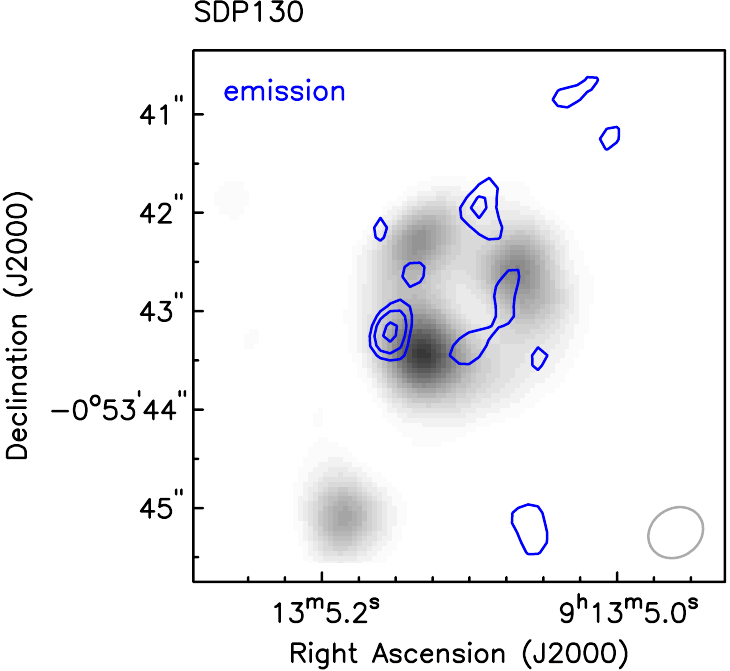}
 \caption{As in Extended Data Fig.2, but for NAv1.56, NAv1.144 and SDP130.
  Only emission is detected in SDP130. } 
\end{figure}

\end{document}